\documentclass[onecolumn,twoside,12pt,romanappendices]{IEEEtran}
\usepackage{setspace}
\usepackage{cite,color}%
\usepackage{amsfonts}%
\usepackage{amsmath,url}%
\usepackage{amssymb}%
\usepackage{graphicx}
\usepackage{balance,epsfig}
\usepackage{amsmath,amssymb,graphicx,epsfig,url,cite,balance}
\begin{document}
\newcounter{eqncount}
\pagestyle{empty}
\thispagestyle{empty}

\title{Level Crossing Rates of Interference in Cognitive Radio Networks}
\author{\authorblockA{Muhammad Fainan Hanif and Peter J.
Smith}\\
\authorblockA{Department of Electrical and Computer Engineering, University of Canterbury,
Christchurch, New Zealand}\\
Email: mfh21@student.canterbury.ac.nz,~p.smith@elec.canterbury.ac.nz
\footnote{Part of this work was presented at the IEEE AusCTW
\cite{AusCTW} in Sydney in Feb. 2009.}} \maketitle
\thispagestyle{empty} \pagestyle{empty}
\begin{abstract}
The future deployment of cognitive radios (CRs) is critically
dependent on the fact that the incumbent primary user (PU) system
must remain as oblivious as possible to their presence. This in turn
heavily relies on the fluctuations of the interfering CR signals. In
this letter we compute the level crossing rates (LCRs) of the
cumulative interference created by the CRs. We derive analytical
formulae for the LCRs in Rayleigh and Rician fast fading conditions.
We approximate Rayleigh and Rician LCRs using fluctuation rates of
gamma and scaled noncentral $\chi^2$ processes respectively. The
analytical results and the approximations used in their derivations
are verified by Monte Carlo simulations and the analysis is applied
to a particular CR allocation strategy.
\end{abstract}
\section{Introduction}
It is now well known \cite{report1,report2} that granting exclusive
licences to service providers for particular frequency bands has
resulted in severe under-utilization of the radio frequency (RF)
spectrum. This has led to global interest in the concept of
cognitive radios (CRs) or secondary users (SUs). These CRs are
deemed to be intelligent agents capable of making opportunistic use
of radio spectrum while simultaneously existing with the legacy
primary users (PUs) without harming their operation.

In addition to ensuring their own quality of service (QoS)
operation, the most important and challenging task for the CRs is to
avoid adverse interference to the incumbent PUs. Hence, it is
necessary to develop schemes that can help PUs avoid such harmful
interference. The recently developed \cite{ICC,AusCTW} methods based
on radio environment maps (REMs) \cite{J.Reed1,J.Reed2} can help
achieve this goal very efficiently. In \cite{ICC} only those CRs are
allowed to operate in a particular time, frequency or space slot
that do not reduce the PU signal to noise ratio (SNR) by more than
some agreed penalty. However, the approaches in \cite{ICC,AusCTW}
allow CRs to operate on the basis of average signal to interference
plus noise (SINR) ratio values and do not consider the instantaneous
temporal variation of the interference. Throughout the paper SNR or
SINR values represent long term average values while the
interference is considered on an instantaneous scale. Note that
small scale variations in the composite CR interference can degrade
the PU performance even though the CR levels may be acceptable on
average. Thus, the determination of the rate at which the
instantaneous interference crosses a particular threshold and the
duration for which it stays above or below it, is an issue of core
importance. For PU system designers, the following questions are
important:
\begin{itemize}
\item How large is the CR-PU interference and can it be controlled?
\item How often will the interference exceed a threshold?
\item How long does the interference stay above a given
threshold?
\item How do these issues vary with the type of fading?
\end{itemize}
These questions form the focus of this paper. In particular, we make
the following contributions:
\begin{itemize}
\item We determine the level crossing rate (LCR) and average exceedance
duration (AED) of the CR-PU interference for Rayleigh and Rician
fading channels, and various CR interferer profiles.
\item For Rayleigh channels, we approximate the LCRs using
fluctuation rates of a gamma process. Similarly, for Rician fading
we approximate the instantaneous aggregate interference with a
fractional order noncentral $\chi^2$ variable to evaluate the LCRs.
These approximations are validated via simulations.
\item For CR systems where the long term interference has an imposed maximum, results
show that the LCR is maximum at or around the maximum interference
threshold and is virtually zero 5 dB beyond this point. We also show
that compared to Rayleigh fading, in line of sight (LOS) channels,
the interference rarely crosses the threshold and when it does, it
only exceeds the threshold value for a short duration.
\end{itemize}
The rest of the paper is organized as follows: Section~II
characterizes the instantaneous interference and derives the LCR and
AED results. In Section~III we present simulation and analytical
results. Finally, in Section~IV we describe our conclusions.
\section{Instantaneous CR performance}
In any CR allocation policy, for example \cite{ICC}, even if the
target SINR of the PU is exactly met the fast fading will result in
fluctuations of the instantaneous SINR both above and below the
target. As a first look at this problem we fix the PU signal power
and consider the instantaneous variation of the interference only.
Hence, in this section we focus on the instantaneous temporal
behavior of the aggregate interference. For this purpose we evaluate
the LCR (and thus the average exceedance duration (AED)) of the
cumulative interference offered by the CRs. First we calculate the
LCRs for a Rayleigh environment and then we characterize them for
Rician fading conditions.
\subsection{LCRs for Rayleigh Fading}
For a given set of CR interferers, the instantaneous aggregate
interference, $I_{Ray}(t)$, is given by:
\begin{equation}\label{instRay}
I_{Ray}(t)=\sum_{i=1}^NI_i|h_i(t)|^2
\end{equation}
where $I_i$ represents the long term interference power of the $i$th
CR, $h_i(t)$ is the corresponding normalized channel gain so that in
Rayleigh fading $|h_i(t)|^2$ is a standard exponential random
variable with unit mean and $N$ is the number of interfering CRs.
Note that we fix the long term interference values, $I_1,\ldots,I_N$
and consider the variation of the fast fading terms, $h_i$. From
(\ref{instRay}), the aggregate interference is represented as a
weighted sum of exponential variables. Such weighted sums can be
approximated by a gamma variable \cite{Kutz}. Simulated results show
that the gamma fit is very good, but are not shown here for reasons
of space. However, the corresponding LCR results are shown to be
accurate in Figs.~\ref{fig2}-\ref{fig4}. Note that the exact LCR
computation for such sums was given in \cite{MRC2} for the case of
three and four branch maximal ratio combining (MRC) by providing
special function integrals. Recently, more general expressions for
arbitrary number of branches have been derived in \cite{MRC1}.
However, the approach of \cite{MRC1} results in numerical
difficulties, especially for large values of $N$, which can be the
case for CR systems. Hence, an approximation is useful to overcome
these problems and to provide a much simpler solution. Thus,
approximate LCRs for (\ref{instRay}) can be found by calculating the
LCR of the equivalent gamma process.
The LCR for a gamma process has been calculated in \cite{lcrGamma}.
Therefore, the crossing rate of $I_{Ray}(t)$ across a threshold,
$T$, can be approximated by:
\begin{equation}\label{lcrRay}
\textrm{LCR}_{I_{Ray}}(T)=\frac{1}{2\Gamma(r)}\sqrt{\frac{2|\ddot{R}(0)|}{\pi}}(\theta
T)^{r-0.5}\exp(-\theta T)
\end{equation}
where $r=E(I_{Ray}(t))^2/Var(I_{Ray}(t))$,
$\theta=E(I_{Ray}(t))/Var(I_{Ray}(t))$ and
$\ddot{R}(0)=\ddot{\rho}_{Ray}(0)$ is the second derivative of the
autocorrelation function (ACF) of $I_{Ray}(t)$ at time lag,
$\tau=0$. Hence, to compute the LCR in (\ref{lcrRay}) only the mean,
variance and ACF of the random process in (\ref{instRay}) are
required.
The first two moments of (\ref{instRay}) can be computed as
$E(I_{Ray}(t))=\sum_{i=1}^NI_i$ and
$Var(I_{Ray}(t))=\sum_{i=1}^NI_i^2$. To calculate the ACF, note
that:
\begin{equation}\label{corrEq}
h_i(t+\tau)=\rho_i(\tau)h_i(t)+\sqrt{(1-\rho_i^2(\tau))}e_i(t),
\end{equation}
where $e_i(t)$ is independent of $h_i(t)$ and statistically
identical to $h_i(t)$. Assuming a Jakes' fading process,
$\rho_i(\tau)$ is the zeroth order Bessel function of the first
kind, $J_0(2\pi f_D\tau)$ and $f_D$ is the Doppler frequency. Using
(\ref{corrEq}) we have:
\setlength{\arraycolsep}{0.0em}
\begin{eqnarray}\label{meantT}
E[I_{Ray}(t)I_{Ray}(t+\tau)]&{}={}&\sum_{i,j=1}^NI_iI_jE[|h_i(t)|^2|h_j(t+\tau)|^2]\nonumber\\
&{}={}&\sum_{i\neq
j}^NI_iI_j+\bigg(\sum_{i=1}^NI_i^2E[|h_i(t)|^2(\rho_i^2(\tau)
{\times}|h_i(t)|^2+(1-\rho_i^2(\tau))|e_i(t)|^2)]\bigg)\nonumber\\
&{}={}&\sum_{i\neq j}^NI_iI_j+\sum_{i=1}^NI_i^2+\sum_{i=1}^NI_i^2\rho_i^2(\tau)\nonumber\\
&{}={}&\bigg(\sum_{i=1}^NI_i\bigg)^2+\sum_{i=1}^NI_i^2\rho_i^2(\tau),
\end{eqnarray}
\setlength{\arraycolsep}{5pt}
\hspace{-1.5mm}where in the second to last step above, we have used
the fact that cross products have zero mean and that
$E[|h_i(t)|^4]=2$. The ACF of (\ref{instRay}) is given by:
\begin{equation}\label{acf}
\rho_{Ray}(\tau)\!=\!\frac{E(I_{Ray}(t)I_{Ray}(t+\tau))\!-\!E(I_{Ray}(t))E(I_{Ray}(t+\tau))}{\sqrt{Var(I_{Ray}(t))Var(I_{Ray}(t+\tau))}},
\end{equation}
and with the relevant substitutions, the ACF becomes:
\begin{equation}\label{trueAcf}
\rho_{Ray}(\tau)=\frac{\sum_{i=1}^NI_i^2J_0^2(2\pi
f_D\tau)}{\sum_{i=1}^NI_i^2}.
\end{equation}
Finally, using the expansion $J_0(2\pi
f_D\tau)=1-\pi^2f_D^2\tau^2+\ldots$, the second derivative of the
ACF needed to compute the LCR in (\ref{lcrRay}) is evaluated as:
\begin{equation}\label{dder}
\ddot{\rho}_{Ray}(0)=-4\pi^2\frac{\sum_{i=1}^NI_i^2f_D^2}{\sum_{i=1}^NI_i^2}.
\end{equation}
Hence, the three parameters, $r$, $\theta$ and $\ddot{R}(0)$, are
available and (\ref{lcrRay}) gives the approximate LCR.
\subsection{LCRs for Rician Fading}
The instantaneous aggregate interference, $I_{Ric}(t)$, for this
scenario is given as:
\begin{equation}\label{instRic}
I_{Ric}(t)=\sum_{i=1}^NI_i|h_i(t)|^2,
\end{equation}
where $|h_i(t)|$ is Rician, with Rician K-factor denoted by $K$, and
$N,I_1,I_2,\ldots,I_N$ are as defined in (\ref{instRay}). Hence,
$I_{Ric}(t)$ is a weighted sum of noncentral chi-square ($\chi^2$)
random variables. Note that standard LCR results for Ricians
\cite{Stuber,LCR_NC2} and noncentral $\chi^2$ variables
\cite{LCR_NC2} cannot be applied directly here. The work in
\cite{patzold} is for a single Rician and in \cite{LCR_NC2} the LCR
applies to the case where $I_1=I_2=\ldots=I_N$ and an exact
noncentral $\chi^2$ arises with integer degrees of freedom (dof).
Instead, using the same approximation philosophy as that used in the
Rayleigh case, we propose approximating (\ref{instRic}) by a single
noncentral $\chi^2$. This approach is less well documented but has
appeared in the literature (see \cite{Kim}). Also note that a
scaled, rather than a standard, noncentral $\chi^2$ distribution is
required for fitting and the resulting best-fitting distribution
will almost certainly not have integer dof. A noncentral $\chi^2$
variable with $v$ dof, non-centrality parameter $\lambda$ and scale
parameter $\alpha$ has the following PDF:
\begin{equation}\label{ncx2}
p(x)=\frac{\alpha}{2}\exp\bigg(\frac{-(\lambda+\alpha
x)}{2}\bigg)\bigg(\frac{\alpha
x}{\lambda}\bigg)^{\frac{v-2}{4}}I_{\frac{v-2}{2}}\big(\sqrt{\lambda
\alpha x}\big),
\end{equation}
where $I_{(v-2)/2}$ is a modified Bessel function of the first kind
with order $(v-2)/2$. Fitting the PDF in (\ref{ncx2}) to the
variable in (\ref{instRic}) is performed using the method of moments
technique so that the approximate noncentral $\chi^2$ has the same
first three moments as $I_{Ric}(t)$. The derivation details are
outlined in Appendix I. Note that there can be numerical
difficulties with the approach for certain values of
$I_1,I_2,\ldots,I_N$. However, when this approach does not work it
is straightforward to perform a numerical minimization of the
difference between the true moments of the CR interference and the
moments of the $\alpha-$scaled noncentral $\chi^2$ variable. Values
of $\lambda$, $v$ and $\alpha$ which minimize this difference can
then be used.

The LCR of a noncentral $\chi^2$ process with integer dof can be
readily obtained from \cite{LCR_NC2}. In particular, if we
substitute $R=T$, $\sigma^2=1$, $M=v/2$, $s^2=\lambda$ and $f_m=f_D$
in \cite[Eq. (15)]{LCR_NC2} we get the following expression for the
LCR of the $\alpha-$scaled noncentral $\chi^2$ variable
\setlength{\arraycolsep}{0.0em}
\begin{eqnarray}\label{lcrRic1}
\textrm{LCR}_{I_{Ric}}&{}={}&\sqrt{\pi}f_D(\alpha
T)^{\frac{v}{4}}\lambda^{\frac{-(v-2)}{4}}e^{\big(\frac{-\lambda-\alpha
T}{2}\big)}I_{\frac{v-2}{2}}\big(\sqrt{\lambda\alpha T}\big).
\end{eqnarray}
\setlength{\arraycolsep}{5pt}
The result in (\ref{lcrRic1}) holds good for a noncentral $\chi^2$
process with integer dof. In Appendix II we show that this formula
is also valid for non-integer dof. Note that a similar extension for
a central $\chi^2$ with integer order \cite{chiP} to a central
$\chi^2$ with fractional order \cite{lcrGamma} has also been shown
to be correct.
\subsection{AEDs}
We define the AED for both Rayleigh and Rician environments as the
average time that the aggregate interference stays above a given
threshold $T$ \cite{Stuber}. Mathematically,
\begin{equation}\label{aed}
AED=\frac{1-F(T)}{LCR}
\end{equation}
where $F(T)$ gives the distribution function of the aggregate
interference. Note that the exact CDFs of both $I_{Ray}(t)$ and
$I_{Ric}(t)$ can be found in \cite{Kutz-old}.
\section{Results}
In order to evaluate the accuracy of (\ref{lcrRay}) and
(\ref{lcrRic1}) it is important to use realistic values of
$I_1,I_2,\dots,I_N$. Hence, we use a particular CR access scheme
\cite{ICC} to provide these values. The \emph{decentralized
selection} algorithm in \cite{ICC} employs a controller that
considers CRs in their order of arrival. Each interferer is
considered in turn and is accepted if the combined interference from
previously accepted CRs and the current CR is less than some
interference threshold. If a CR is not accepted, the next CR in the
list is investigated. The $I_i$ values are generated in \cite{ICC}
from randomly located CRs in a circular region and include path loss
and shadowing effects. In \cite{ICC} a threshold value is used which
corresponds to the PU accepting a $2$ dB loss in its SNR due to the
presence of CRs.

From 1000 simulations, using the above selection procedure two sets
of interferers were selected. The first set selected had the highest
variance. Only 3 CRs were accepted and there was a dominant
interferer which accounted for 95\% of the interference power. The
second set had the lowest variance, representing the no dominant
interferer scenario. Here, 18 CRs were accepted with the largest
interferer only accounting for 16\%. In addition to giving examples
of engineering importance (presence or absence of a dominant
interferer) these two cases also test the general applicability of
(\ref{lcrRay}) and (\ref{lcrRic1}) over a wide range of interferer
profiles. These sets were obtained using the following parameter
values: shadow fading variance, $\sigma=8.0$ dB, path loss exponent,
$\gamma=3.5$, radius of PU coverage area, $R=1000$ m, radius of CR
coverage area, $R_c=100$ m, CR density of 1000 CRs per square
kilometer, an activity factor of $0.1$ and $f_D=25$ Hz.
\subsection*{LCR and AED of CR-PU Interference}
\begin{figure}[t]
\centering
\includegraphics[width=0.65\columnwidth]{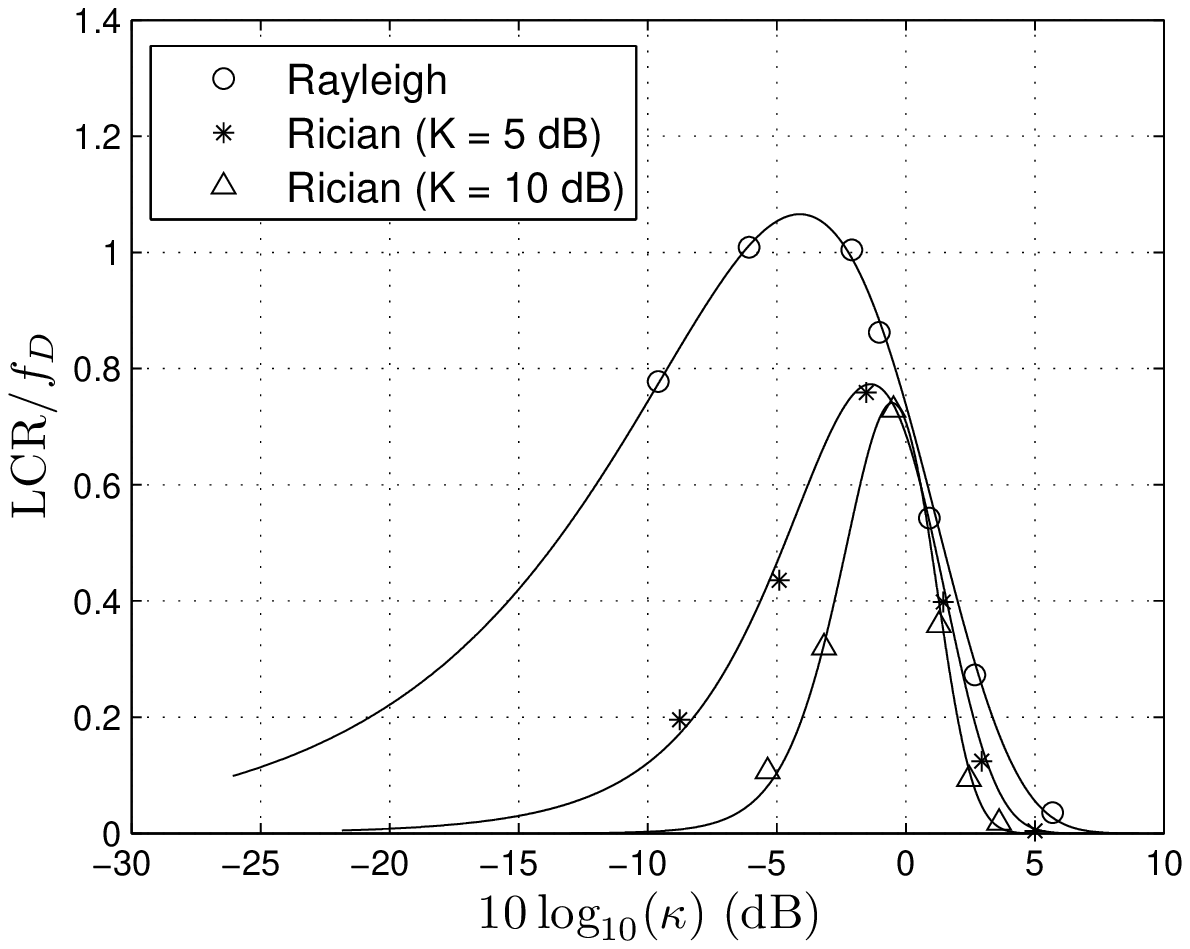}
\caption{LCR results for different fading conditions with dominant
interferers. The solid lines represent analytical results.
Simulation values are shown by the circle, star and triangle
symbols.} \label{fig2}
\end{figure}

\begin{figure}[t]
\centering
\includegraphics[width=0.65\columnwidth]{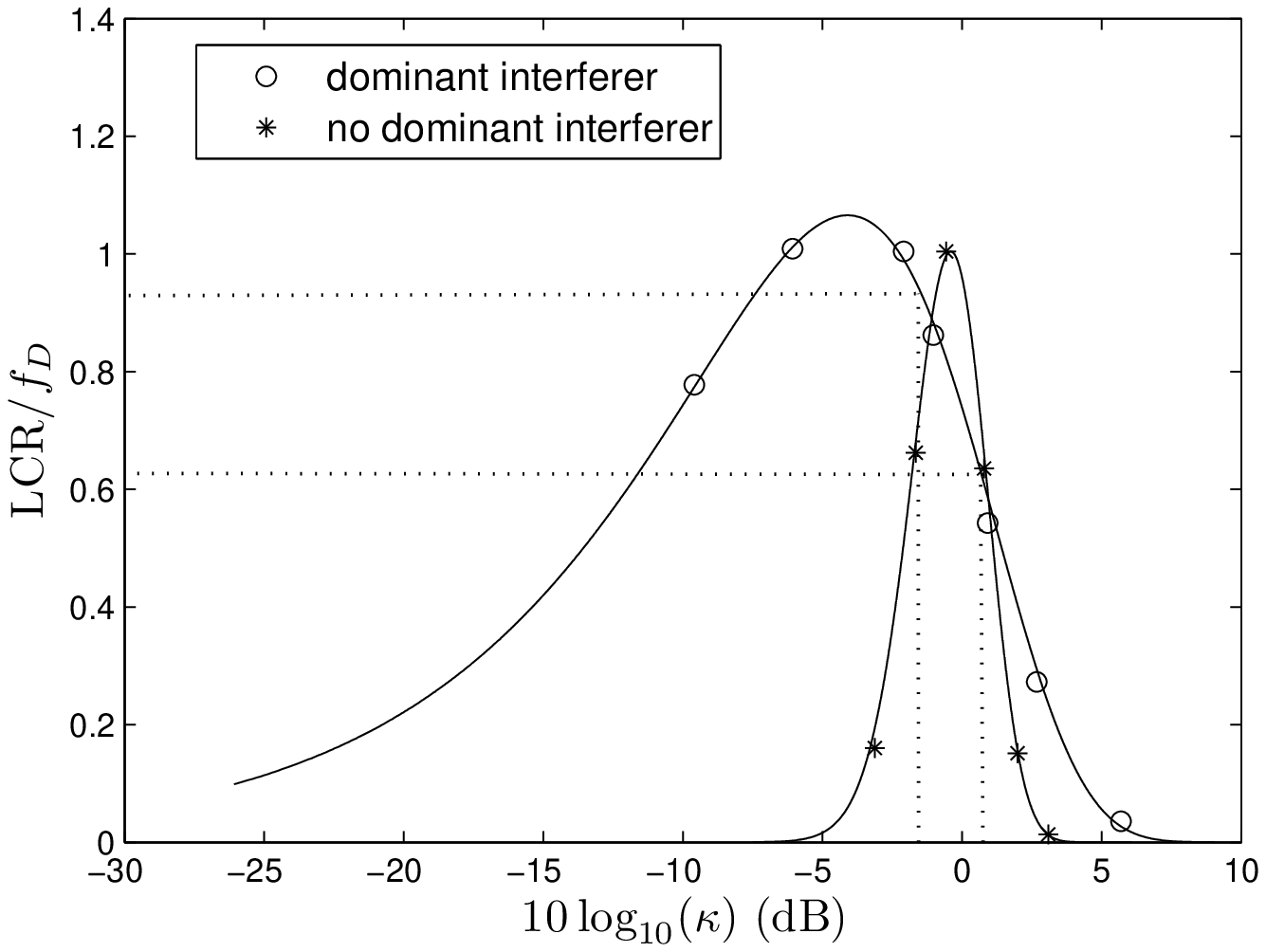}
\caption{LCR results for the dominant and no dominant interferer
cases in a Rayleigh fading scenario. The solid lines represent
analytical results. Simulation values are shown by the circle and
star symbols. The interference threshold values and their LCRs are
shown by dotted lines.} \label{fig3}
\end{figure}

\begin{figure}[t]
\vspace{0.0mm} \centering
\includegraphics[width=0.65\columnwidth]{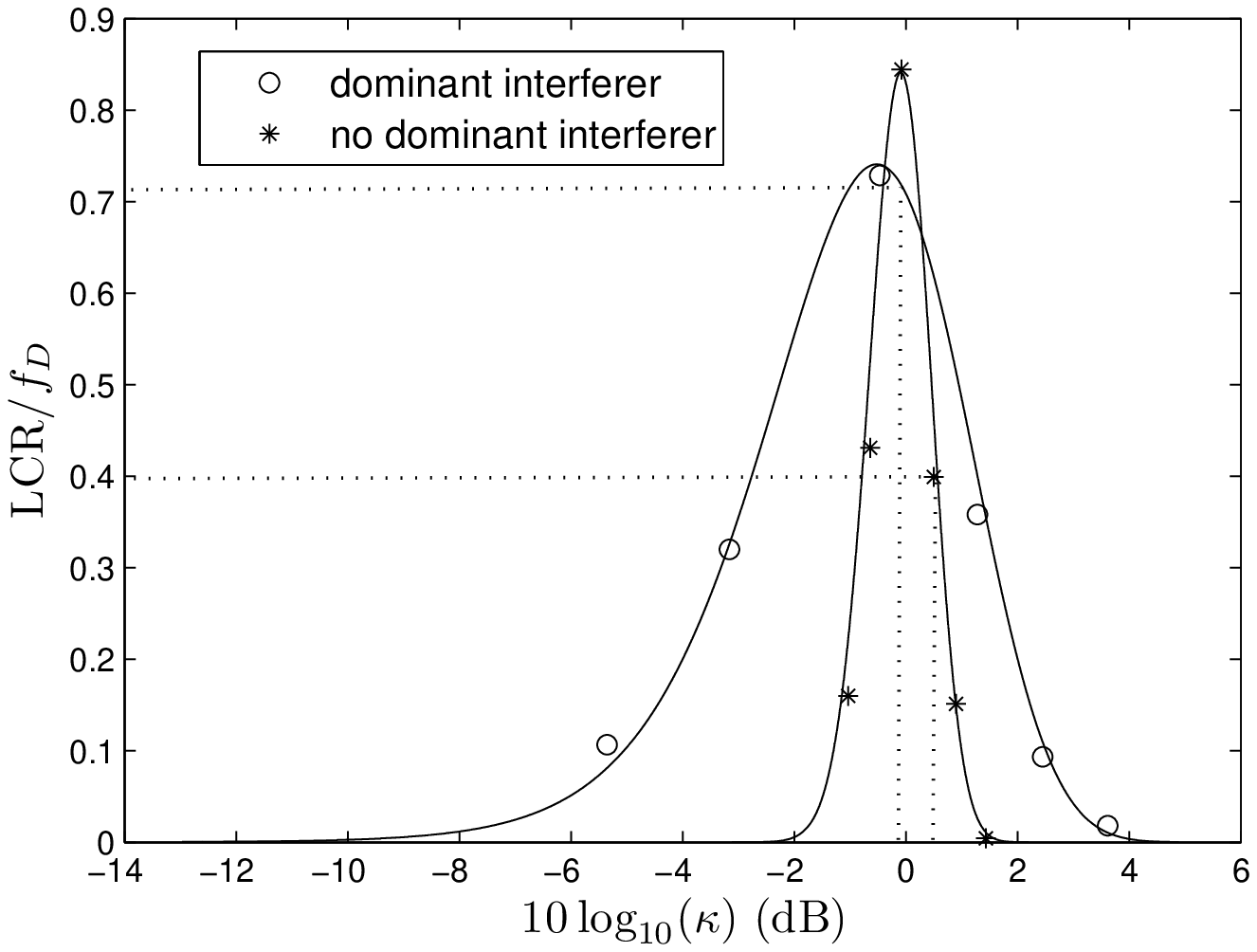}
\caption{LCR results for the dominant and no dominant interferer
cases in a Rician ($K=10$ dB) fading scenario. The solid lines
represent analytical results. Simulation values are shown by the
circle and star symbols. The interference threshold values and their
LCRs are shown by dotted lines.} \label{fig4}
\end{figure}

\begin{figure}[t]
\centering
\includegraphics[width=0.65\columnwidth]{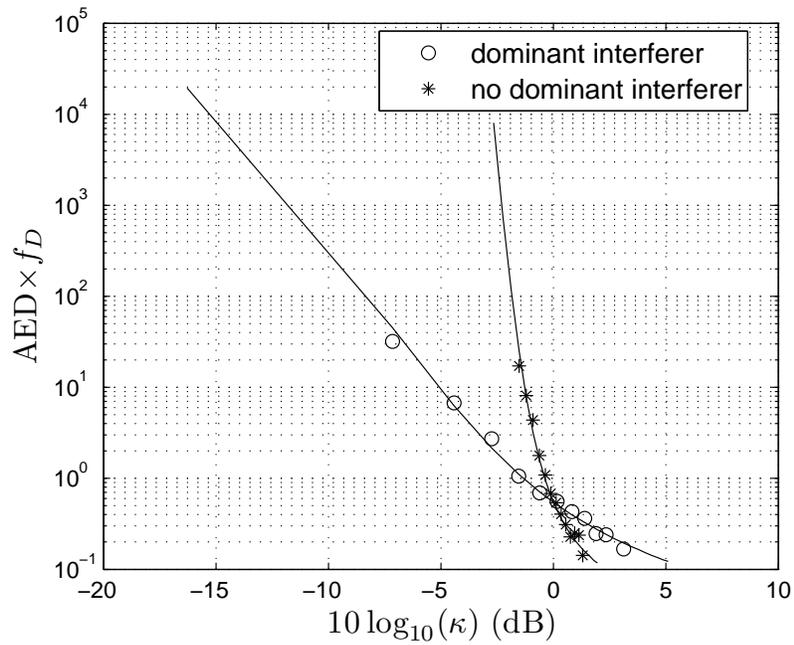}
\caption{AED results for the dominant and no dominant interferer
cases in a Rician ($K=10$ dB) fading scenario. The solid lines
represent analytical results. Simulation values are shown by the
circle and star symbols.} \label{fig_1} \vspace{-3mm}
\end{figure}

Figures \ref{fig2}, \ref{fig3} and \ref{fig4} show the LCR
(normalized by Doppler frequency) of the interference for different
types of fading and interference profiles. The $x$-axis is also
normalized by the rms value of the process so that
$\kappa=T/\sqrt{m_2}$ is plotted, where $T$ is the interference
power level and $m_2$ is the mean-square interference (see Appendix
I). Figure \ref{fig2} shows the effect on LCR of increasing the
Rician $K$-factor, with the strong LOS case being considerably
narrower than the non-LOS case. Figures \ref{fig3} and \ref{fig4}
also show the value of the normalized interference threshold that
restricts the long term average interference value in the CR system
(as shown by the dotted lines). Note that there are multiple
thresholds since the normalization is different for different
channels. For all types of fading, the maximum LCR is observed close
to this threshold value. This is because the CR allocation method
gives a mean interference level close to the threshold. Even in
strong LOS conditions ($K=10$ dB), the interference shows a
significant number of level crossings across the buffer due to the
scattered component. Figure~\ref{fig3} shows the case of Rayleigh
fading where the interference budget is dominated by a single large
interferer with a number of smaller additional interferers. Also
shown is the no dominant interferer case. Figure~\ref{fig4} shows
the same results for a Rician channel with $K=10$ dB.
Figures~\ref{fig3} and \ref{fig4} show that when there are many
small interferers, the interference is more stable compared to the
dominant interferer case. The results in Fig.~\ref{fig4} are quite
promising. In near LOS conditions, the interference has a much lower
level crossing rate across the interference buffer for the no
dominant interferer case. Hence, it may be a desirable part of the
CR allocation policy to avoid any single user which takes up a
significant part of the buffer. Finally, for completeness, we show
the AED results corresponding to Fig.~\ref{fig4} in
Fig.~\ref{fig_1}. As expected, the time spent by the interference
above a threshold decreases as the threshold value increases.
Therefore, for the no dominant interferer case, the interference
seldom crosses the threshold (see Fig. \ref{fig4}), and when it
does, it only exceeds the threshold for a small period of time.
Finally, we note that all figures show an excellent agreement
between the analytical approximations and the simulations.
\section{Conclusion}
In this paper we determine the LCR and AED for the CR-PU
interference for Rayleigh and Rician channels. We have shown that
LCRs in Rayleigh environment can be accurately approximated by LCRs
of a gamma process. Similarly, while deriving LCR approximations in
Rician conditions we have shown that the LCR of a noncentral
$\chi^2$ process with non-integer dof has the same form as that of a
noncentral $\chi^2$ process with integer dof. The LCR results show
that it is desirable for the interference to be made up of several
small interfering CRs rather than a dominant source of interference.
The LCR of the former case is more stable than the latter. The AED
results also show that the interference exceeds the threshold value
for small periods of time in the latter case.
\appendices
\section{}
Let $Y$ denote the random variable defined by (\ref{ncx2}). The
first three moments of $Y$ are \cite{Kutz}:
\begin{equation}\label{app1}
E(Y)={\alpha^{-1}}{(\lambda+\nu)}
\end{equation}
\begin{equation}\label{app2}
E(Y^2)= {\alpha^{-2}} {((\lambda+\nu)^2+2(\lambda+\nu)+2\lambda)}
\end{equation}
\begin{equation}\label{app3}
E(Y^3)={\alpha^{-3}}({(\lambda+\nu)^3+6(\lambda+\nu)^2+2\lambda(\lambda+\nu)+8(\lambda+\nu)+16\lambda}).
\end{equation}
Similarly, suppose $m_1, m_2$ and $m_3$ denote the moments of
$I_{Ric}(t)$ in (\ref{instRic}) about origin. Expanding
$I_{Ric}(t)$, $I_{Ric}^2(t)$ and $I_{Ric}^3(t)$ into multiple sums
and taking expectation using standard results in \cite{Kutz} leads
to:
\begin{equation}\label{app4}
m_1=\sum_{i=1}^NI_i
\end{equation}
\begin{equation}\label{app5}
m_2=\bigg(\sum_{i=1}^NI_i\bigg)^2+\sum_{i=1}^NI_i^2\bigg(1-\bigg(\frac{K}{K+1}\bigg)^2\bigg)
\end{equation}
\begin{equation}\label{app6}
m_3=\bigg(\sum_{i=1}^NI_i\bigg)^3+3\sum_{i=1}^N\sum_{k\neq
i,k=1}^NI_i^2I_k\bigg(1-\bigg(\frac{K}{K+1}\bigg)^2\bigg)+\sum_{i=1}^NI_i^3\bigg(2-6\bigg(\frac{K}{K+1}\bigg)^2+4\bigg(\frac{K}{K+1}\bigg)^4\bigg).
\end{equation}
Now applying the method of moments, we solve $m_k=E(Y^k)$ for
$k=1,2,3$ and obtain the following:
\begin{equation}\label{app7}
\lambda=0.5\alpha(\alpha m_2-\alpha m_1^2-2m_1),\quad \nu=\alpha
m_1-\lambda,
\end{equation}
%\begin{equation}\label{app8}
%\nu=\alpha m_1-\lambda,
%\end{equation}
where $\alpha$ is the solution to the following quadratic equation:
\begin{equation}\label{app9}
\alpha^2 m_1^3-\alpha^2 m_3+6\alpha m_1^2+3\alpha^2
m_1m_2-3\alpha^2m_1^3+8\alpha m_2-8\alpha m_1^2-8m_1=0.
\end{equation}
\section{}
In \cite{LCR_NC2} a noncentral $\chi^2$ process, denoted $r$ is
considered. The only part of the derivation in \cite{LCR_NC2} that
requires integer dof is the proof that the conditional distribution
of $\dot{r}$ given $r$ is Gaussian with variance
$Var(\dot{r}|r)=4\tilde{\sigma}^2r$ where $\tilde{\sigma}^2$ is a
variance parameter. In this Appendix we show that this is true for a
general noncentral $\chi^2$ process. The LCR of a stationary gamma
process was first derived by Barakat \cite{lcrGamma} in an optics
context building on previous results in \cite{Barakat}. This
analysis is based on the representation \cite{lcrGamma}
\begin{equation}
\Omega=\int_{A}|E(x)|^2dx
\end{equation}\label{appb1}
where $A$ is the region of integration (an aperture in
\cite{lcrGamma}), $E(x)$ is a circular complex zero-mean Gaussian
process and $\Omega$ is the resulting gamma variable. If $E(x)$ is
allowed to have a constant non-zero mean then for certain $A$, the
resulting $\Omega$ has a scaled noncentral $\chi^2$ distribution
with arbitrary degrees of freedom (not necessarily integer). For
this noncentral case let $E(x)=E_1(x)+jE_2(x)$ and
\begin{equation}
\Omega=\int_{A}E_1^2(x)+E_2^2(x)dx,
\end{equation}\label{appb2}
where $E_1(x)$ and $E_2(x)$ are both non-zero mean Gaussian
processes. The derivative of $\Omega$ is therefore
\begin{equation}
\dot{\Omega}=\int_{A}[2E_1(x)\dot{E_1}(x)+2E_2(x)\dot{E_2}(x)]dx.
\end{equation}\label{appb3}
Now it is well known \cite{Rice,LCR_NC2,MRC1,MRC2,patzold} that
$\dot{E_1}(x)$, $\dot{E_2}(x)$ are zero-mean Gaussian variables
which are independent of $E_1(x)$, $E_2(x)$ and each other. Let the
distribution of both derivatives be denoted by
$\mathcal{N}(0,\sigma^2)$. Hence, conditioned on $E_1(x)$ and
$E_2(x)$ over $x\in A$, the derivative, $\dot{\Omega}$, is also zero
mean Gaussian. The variance of $\dot{\Omega}$ conditioned on
$\{E_1(x),E_2(x)|x\in A\}$ is given by
\setlength{\arraycolsep}{0.0em}
\begin{eqnarray}\label{appb4}
&{}E{}&\bigg[\int_A\int_A(2E_1(x)\dot{E_1}(x)+2E_2(x)\dot{E_2}(x))(2E_1(y)\dot{E_1}(y)+2E_2(y)\dot{E_2}(y))dxdy\bigg]\nonumber\\
&{}={}&E\bigg[\int_A\big(4E_1^2(x)\dot{E_1^2}(x)+4E_2^2(x)\dot{E_2^2}(x)\big)dx\bigg]
\end{eqnarray}
\setlength{\arraycolsep}{5pt}
since $\dot{E_i}(x)$ is independent of $\dot{E_i}(y)$ for $x\neq y$.
Also, since $E[\dot{E_i}^2(x)]=\sigma^2$, the conditional variance
is
\begin{equation}
4\sigma^2\int_A [E_1^2(x)+E_2^2(x)]dx=4\sigma^2\Omega
\end{equation}\label{appb5}
Hence, $\dot{\Omega}$ has the representation
$\dot{\Omega}=2\sigma\Omega^{1/2}Z$ where $Z\sim \mathcal{N}(0,1)$
and $\dot{\Omega}$ has the conditional density
\begin{equation}
f_{\dot{\Omega}|\Omega}=\frac{\exp(\frac{-x^2}{8\sigma^2\Omega})}{\sqrt{8\pi\sigma^2\Omega}}.
\end{equation}\label{appb6}
Since $\dot{\Omega}\sim\mathcal{N}(0,4\sigma^2\Omega)$, conditional
on $\Omega$, the proof is complete.
\section*{Acknowledgment}
The authors wish to acknowledge the financial support provided by
Telecom New Zealand and National ICT Innovation Institute New
Zealand (NZi3) during the course of this research.
\balance
%\nocite{*}
\bibliographystyle{IEEEtran}
\bibliography{IEEEabrv,AusCTW}
\newpage
\end{document}